\documentstyle[12pt]{article}
\begin{document}
\thispagestyle{empty}
\begin{center}{{Dynamo waves in Friedmann and Misner cosmologies with torsion as generators of magnetic fields in galactic clusters}}
\end{center}
\vspace{1.0cm}
\begin{center}
{\large By L.C. Garcia de Andrade\footnote{Departamento de
F\'{\i}sica Te\'{o}rica - IF - UERJ - Rua S\~{a}o Francisco Xavier
524, Rio de Janeiro, RJ, Maracan\~{a}, CEP:20550.
e-mail:garcia@dft.if.uerj.br}}
\end{center}
\begin{abstract}
Earlier a generalised dynamo equation in first order torsion [Phys Lett B 711, (2012) 143] was derived. From this equation it is shown that, two cosmological solutions can be obtained. In both solutions torsion decouples from Einstein gravity and general relativistic effects lead to hyperbolic Friedmann universe in the first case. The main physical result of the paper is simply a Misner solution where torsion is not decoupled from GR but indicates that the intergalactic magnetic field driven from torsion is given by $B_{ICM}\sim{100{\mu}G}$ which is a dynamo wave solution. This is not in agreement with the common sense that dynamo action is not present in the intergalactic space. It is also shown that at $1Mpc$ scales a seed field of $10^{-13}G$ can feed galactic dynamos through dynamo waves. \end{abstract}

\newpage
\section{Introduction}
Earlier Bamba et al \cite{1} addressed the problem of the
amplification of large scale magnetic fields at $1Mpc$ without dynamo action given by $10^{-9}G$ in the galactic intracluster in the special type of torsion theory called teleparallelism \cite{2}. In this paper we show that by making use of a dynamo equation with torsion \cite{3} this time taking into account the expansion of the universe, it is possible to compute the intracluster matter {ICM) using dynamo waves. In $10 pc$ coherence length of the magnetic field and its strength may achieve $B_{ICM}\sim{10^{-4}G}$ as a torsion driven magnetic field which is quite close to the intergalactic magnetic fields of the order $10^{-5}G$ \cite{4}. This may indicate that dynamo waves can allow dynamo action be useful to IGM and not only Biermann battery for example. Kulsrud et al \cite{5}have considered that no further dynamo action can be necessary after galaxy formation. It is also shown that when the magnetic seed field is of the order of  $B_{seed}\sim{10^{-13}G}$ this field can be amplified by dynamo waves to the galactic magnetic fields of strength of $B_{G}\sim{1{\mu}G}$ at scales of $1Mpc$. When dynamo waves are not present the Riemannian or general relativistic and non-Riemannian or torsioned effects are decoupled and the Friedmann universe is found to be hyperbolic.  This shows that dynamo waves can be an important ingredient in the theory of CMF to seed galactic dynamos. Therefore it is hown that a non-teleparallel  theory of gravity with metric-torsion scales and dynamo waves can be useful to generate ICM. The paper is organised as follows: In section 2 we present dynamo equation in curvature-torsion scales generalizing Marklund and Clarkson \cite{6} general relativistic dynamo equation to include torsion and examine the decoupling of the dynamo equation without dynamo waves leading to hyperbolic Friedmann cosmology. Section 3 addresses the problem of dynamo waves and the computation of magnetic field in ICM from Misner solution. Section 4 is left for conclusions.
\newpage
\section{Decoupling between curvature and torsion in dynamo equation} From the reference (\ref{3}) Euler-Lagrange equation the curvature  leads to the dynamo equation \cite{7} in 3-vector notation in an expanding universe\begin{equation}
{\partial}_{t}\textbf{B}-\frac{1}{a}[{\nabla}{\times}[\textbf{V}{\times}\textbf{B}]-{\eta}[{\Delta}\textbf{B}-{\nabla}.\textbf{T}{\times}\textbf{B}]]-2\frac{\dot{a}}{a}\textbf{B}=0
\label{1}
\end{equation}
where $T$ here is the actual terrestrial torsion
given in Laemmerzahl \cite{8} given by $T\sim{10^{-17}cm^{-1}}$ or $10^{-31}GeV$. Here $\eta$ is the diffusion or the electric resistivity of the plasma flow and $\textbf{B}$ is the magnetic field. The metric used in the first equation is the Friedmann general metric
\begin{equation}
ds^{2}=dt^{2}-a^{2}(t)[dx^{2}+dy^{2}+dz^{2}]
\label{2}
\end{equation}
Let us now consider the magnetic field in the form
\begin{equation}
B=B_{seed}exp[{\gamma}t-i\textbf{K}\textbf{x}]
\label{3}
\end{equation}
Substitution of this term into generalised curvature-torsion dynamo equation one obtains the uncoupled equations without dynamo waves
\begin{equation}
2\dot{a}+{\gamma}a+{\eta}k^{2}=0
\label{4}
\end{equation}
where ${\gamma}$ is the dynamo amplification factor, and 
\begin{equation}
v\sim{T} \label{5}
\end{equation}
note that from this second equation torsion contributes to enhance the velocity of the plasma. Equation (\ref{4}) has a simple hyperbolic Friedmann solution
\begin{equation}
a(t)=a_{0}sinh[\frac{{\gamma}}{2}t] \label{6}
\end{equation}
Note that in equation (\ref{5}) if we consider above torsion one yields a plasma velocity of $10^{-17} cm.s^{-1}$ for example a non-relativistic electron. This is almost a comoving coordinate. This is an extremely low velocity in the plasma medium . In the next section we shall see that it will be more simple compute the magnetic field when we introduce dynamo waves. 
\section{Dynamo waves in Friedmann universe and ICM}
In this section we shall looking for dynamo waves solutions of dynamo equation in Friedmann universes endowed with torsion. In this case the  GR dynamo equation does not decoupled for the torsion part and the universe left is not hyperbolic anymore. Actually expansion is linear in time and we have a sort of Misner universe. Let us consider then the magnetic field in the form of dynamo waves as
\begin{equation}
B=B_{seed}exp[i[{\gamma}t-\textbf{K}\textbf{x}]]
\label{7}
\end{equation}
substitution of this equation into the dynamo equation yields the equation 
\begin{equation}
i{\gamma}\textbf{B}=\frac{1}{a}i\textbf{k}{\times}(\textbf{v}{\times}\textbf{B})-2\frac{\dot{a}}{a}\textbf{B}+\frac{{\eta}}{a}[k^{2}-i\textbf{k}.\textbf{T}]\textbf{B}
\label{8}
\end{equation}
This dynamo wave equation is also decoupled which yields
\begin{equation}
\dot{a}-\frac{\eta}{2}k^{2}=0
\label{9}
\end{equation}
which yields the following Misner like solution
\begin{equation}
ds^{2}=dt^{2}-\frac{\eta}{2}k^{2}t(dx^{2}+dy^{2}+dz^{2})
\label{10}
\end{equation}
From the torsion equation
\begin{equation}
{\gamma}t=[\frac{v}{\eta}-{k^{-1}T}]
\label{11}
\end{equation}
therefore from this equation one obtains the following cluster torsion induced magnetic field. When the coherent $10kpc$ scale is given we obtain $B_{ICM}\sim{10^{-4}G}$. However if the scale of magnetic fields is larger as $1Mpc$ then the same torsion formula above yields $10^{-6}G$ where the magnetic seed field is taken as $10^{-13}G$ and the torsion above were  used throughout this computation. Therefore one may conclude that this seed field along dynamo waves is enough to seed galactic dynamo. 
\section{Conclusions} Torsion diffusion effects given at galactic disks with the order of ${\eta}\sim{10^{18}cm^{2}s^{-1}}$ for interstellar matter (ISM) along with dynamo waves so commonly used in solar dynamos by Sokoloff \cite{9}. The metric is found from the dynamo equation in curvature-torsion scales, and in the case of non-dynamo waves the Friedmann hyperbolic solution is given. The magnetic fields are then shown to be amplified by dynamo waves. In the case of dynamo waves the Misner metric is found. 
\section{Acknowledgements}
We would like to express my gratitude to D Sokoloff and 
Brandenburg for helpful discussions on the problem of dynamos and
torsion. Special thanks go to A Brandenburg, T Vachaspati and T Kaniahshvilly for their kind hospitality at NORDITA during the COSMOLOGICAL MAGNETIC FIELDS origin and evolution held in Stockholm in the summer of 2015. Financial support University of State of Rio de Janeiro (UERJ) js grateful
acknowledged.
I would like to express my gratitude to Rajeev K Jain and Professor J Yokohama for their many enlightning discussions on the problem of magnetogenesis and dynamo efficiency both at Institute d'Astrophysique at Paris meeting on Primordial Universe after Planck mission held in december 2014. Financial support from University of State of
Rio de Janeiro (UERJ) are grateful acknowledged.

\end{document}